\documentclass[
    ,final            
  ]
  {aipproc}

\layoutstyle{6x9}

\begin{document}

\title{The r-Process in Black Hole Winds}

\classification{26.30.+k; 26.50.+x; 97.10.Tk; 97.60.Bw; 98.35.Bd}
\keywords      {nuclear reactions, nucleosynthesis, abundances
--- black hole physics
--- stars: neutron}

\author{Shinya Wanajo}{
  address={Technische Universit\"at M\"unchen,
        Excellence Cluster Universe,
        Boltzmannstr. 2, D-85748 Garching, Germany;
        shinya.wanajo@universe-cluster.de},
  altaddress={Max-Planck-Institut f\"ur Astrophysik,
        Karl-Schwarzschild-Str. 1, D-85748 Garching, Germany}
}

\author{Hans-Thomas Janka}{
  address={Max-Planck-Institut f\"ur Astrophysik,
        Karl-Schwarzschild-Str. 1, D-85748 Garching, Germany}
}

\begin{abstract}
All the current $r$-process scenarios relevant to core-collapse
 supernovae are facing severe difficulties. In particular, recent
 core-collapse simulations with neutrino transport show no sign of a
 neutron-rich wind from the proto-neutron star. In this paper, we
 discuss nucleosynthesis of the $r$-process in an alternative
 astrophysical site, ``black hole winds'', which are the neutrino-driven
 outflow from the accretion torus around a black hole.  This condition
 is assumed to be realized in double neutron star mergers, neutron star
 -- black hole mergers, or hypernovae.
\end{abstract}

\maketitle


\section{Introduction}

In the past decades, core-collapse supernovae have been considered to be
the most promising astrophysical site that provides the suitable
conditions for nucleosynthesis of the $r$-process. The scenarios include
the neutrino-driven wind \citep{Woos94, Taka94, Qian96, Otsu00, Wana01,
Thom01}, the prompt explosion of a collapsing iron core \citep{Sumi01}
or of an oxygen-neon-magnesium (O-Ne-Mg) core \cite{Wana03}, and the
shocked surface layer of an O-Ne-Mg core \citep{Ning07}. However, recent
hydrodynamical simulations of collapsing iron cores (e.g.,
\citep{Bura06}) and of an collapsing O-Ne-Mg core do not support the
prompt explosion \citep{Kita06} or the shocked surface layer
\citep{Jank08} scenarios. The nucleosynthesis calculations
\citep{Hoff08, Wana09} with these hydrodynamical results also show that
the production of neutron-capture elements proceeds only up to $A =
90$ ($N = 50$). Furthermore, recent long-term simulations of
core-collapse supernovae show that the neutrino-driven outflows are
proton-rich all the way \citep{Fisc09, Hued09}, which poses a severe
difficulty to all the scenarios relevant to the neutrino-driven winds of
core-collapse supernovae.

In contrast, another popular scenario of the astrophysical $r$-process,
the mergers of double neutron stars (NS-NS) \citep{Ruff98} or of a black
hole and a neutron star (BH-NS) \citep{Jank99} in a close binary system
has not been fully explored. The decomposition of cold unshocked
neutron-rich matter from NS-NS is suggested to be an alternative or
additional $r$-process site \citep{Latt77, Meye89, Frei99, Gori05}. In
addition, both NS-NS and BH-NS are expected to form an accretion torus
around a black hole, giving rise to the neutrino-driven winds (``black
hole winds''), which are also expected to provide suitable physical
conditions for the $r$-process \citep{Surm08}.

The reason that the merger scenarios have been disfavored compared to
those of core-collapse supernovae is probably due to discrepancies
between Galactic chemical evolution models and the spectroscopic
analyses of Galactic halo stars. The estimated low event rate ($\sim
10^{-5}$~yr$^{-1}$) of mergers and the long lifetime of the binary
system ($> 100$~Myr) are expected to lead to the delayed appearance of
the $r$-process elements in the Galactic history with too large
star-to-star scattering of their abundances \citep{Arga04}, which is in
conflict with the observational results of halo stars. However, some
recent studies of Galactic chemical evolution based on the hierarchical
clustering of sub-halos \citep{Pran06, Ishi10} do not exclude the
mergers as the dominant astrophysical site of the
$r$-process. Therefore, the mergers cannot be excluded as the
$r$-process site, and more studies of nucleosynthesis are desired when
considering the difficult situation of the supernova scenarios.

In this paper, we examine the $r$-process in black hole winds, which are
common both in NS-NS and BH-NS mergers, and presumably, in ``collapsars''
\citep{MacF99, Prue04}. The previous studies of nucleosynthesis relevant
to these conditions are based on phenomenological models \citep{Prue04,
Surm08}. Currently, however, three-dimensional simulations of the
mergers are out of reach for the wind phase after the formation of a
stable accretion torus \citep{Ruff98, Jank99}. Hence, we apply the
semi-analytic wind model for nucleosynthesis calculations, which has
been developed for the studies of the $r$-process in the neutrino-driven
winds of core-collapse supernovae \citep{Wana01, Wana06}.

\begin{figure}[t]
  \includegraphics[width=0.7\textwidth]{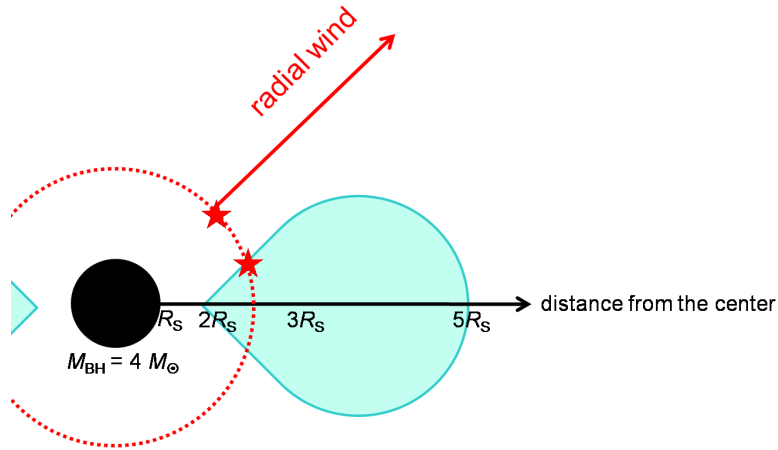}

\caption{Sketch of our model settings for the black hole winds. A
rotating black hole with the mass $M_\mathrm{BH} = 4 M_\odot$ is located in
 the center of an
 accretion torus (``neutrino surface'') that lies between $2
 R_\mathrm{S}$  and $5 R_\mathrm{S}$  from the center, where
 $R_\mathrm{S}$ is the Schwarzschild radius ($= 11.8$~km). The wind is
 assumed to be radial, where the neutrino surface is replaced with an
 equivalent radius from the center (e.g., the star on the dotted circle).}

\end{figure}

\begin{figure}[t]
  \includegraphics[width=0.5\textwidth]{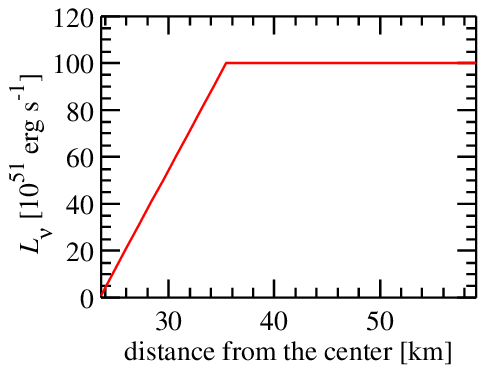}
  \includegraphics[width=0.5\textwidth]{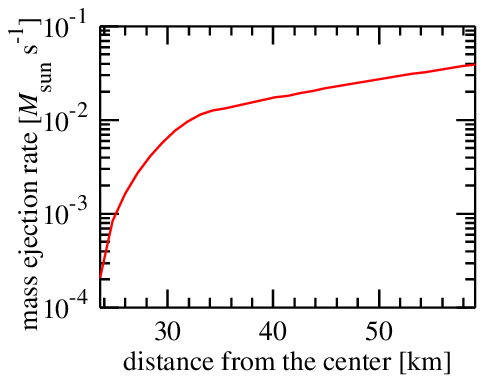}

  \caption{Left: Neutrino luminosity $L_\nu$ as a function of the distance from
 the center. $L_\nu$ is assumed to increase linearly from
 $10^{51}$~erg~s$^{-1}$ to $10^{53}$~erg~s$^{-1}$ between $2
 R_\mathrm{S}$ ($=23.6$~km) and $3 R_\mathrm{S}$ ($=35.4$~km) and take a
 constant value on the outer side. Right: Mass ejection rate $\dot{M}$
 obtained with the $L_\nu$ profile assumed in the left panel, as a
 function of the distance from the center.}
\end{figure}

\section{Modeling the black hole winds}
Our model of black hole winds is based on the semi-analytic, spherically
symmetric, general relativistic model of proto-neutron star winds
\citep{Wana01, Wana06}, as illustrated in Figure~1. The
mass of a central black hole is taken to be $M_\mathrm{BH} = 4 M_\odot$,
which may correspond to, e.g., NS-NS binaries with the equal masses of
$\sim 2 M_\odot$ or BH-NS binaries with the masses of $\sim 2.5 M_\odot$
and $\sim 1.5 M_\odot$. This can be also interpreted as the accreting
black hole of the collapsar from a massive ($> 30 M_\odot$)
progenitor. The accretion torus around the black hole, which is defined
as the ``neutrino surface'', is assumed to lie between $2 R_\mathrm{S}$
($= 23.6$~km) and $5 R_\mathrm{S}$ ($= 35.4$~km) from the center (where
$R_\mathrm{S}$ is the Schwarzschild radius $= 11.8$~km) in the light of
detailed hydrodynamical simulations of BH-NS merging \citep{Jank99}.

In order to connect the aspherical configuration of the winds from the
torus to our spherical model, an arbitrary point on the torus is
replaced by a point on the hypothetical neutrino sphere with an equal
distance from the center, $R_\nu$ (dotted circle in Figure~1). The
solution of the wind from the neutrino sphere with $M_\mathrm{BH}$ and
$R_\nu$ is then obtained in the same manner as for proto-neutron star
winds. The rms average neutrino energies are taken to be 15, 20, and
30~MeV, for electron, anti-electron, and the other flavors of neutrinos,
respectively \citep{Jank99}. The neutrino luminosities of all the
flavors are assumed to be the same value $L_\nu$. The mass ejection rate
at the neutrino sphere $\dot M$ is determined so that the wind becomes
supersonic through the sonic point.

As anticipated from Figure~1, the neutrino flux from the outer regions
of the torus is shielded in the vicinity of the black hole by the
presence of the torus itself. In order to mimic this effect in our
spherical models, we simply assume that $L_\nu$ increases linearly from
$10^{51}$~erg~s$^{-1}$ to $10^{53}$~erg~s$^{-1}$ between $2
R_\mathrm{S}$ ($=23.6$~km) and $3 R_\mathrm{S}$ ($=35.4$~km) and takes a
constant value on the outer side, as shown in Figure~2 (left
panel). This roughly reproduces the peak energy deposition rate by $\nu
\bar{\nu}$ annihilation into $e^+ e^-$ pairs in the vicinity of the
black hole ($\sim 10^{30}$~erg~s$^{-1}$~cm$^{-3}$) \citep{Ruff98,
Jank99}. We define the outflows from $R_\nu < 3 R_\mathrm{S}$ and $R_\nu
> 3 R_\mathrm{S}$ as the inner and outer winds, respectively.

As shown in Figure~2 (right panel), inner winds have rather small
$\dot{M}$ owing to the small $L_\nu$ at $R_\nu$. As a result, the inner
winds obtain substantially higher asymptotic entropies (at 0.5~MeV, up
to $\sim 800 k_\mathrm{B}$ per nucleon, where $k_\mathrm{B}$ is the
Boltzmann constant; Figure~3, left panel) and short expansion timescales
(defined as the $e$-folding time of temperature from 0.5~MeV, down to
$\sim 1$~ms; Figure~3, right panel). This is due to the larger heating
rate \textit{per unit mass} by $\nu \bar{\nu}$ annihilation
\textit{after} leaving the neutrino surface, owing to the smaller matter
density in the inner wind (see the same effect in anisotropic
proto-neutron star winds in \citep{Wana06}). This indicates that the
inner winds are favored for the strong $r$-process (see speculations in
\citep{Ruff98}).

\begin{figure}[t]
  \includegraphics[width=0.5\textwidth]{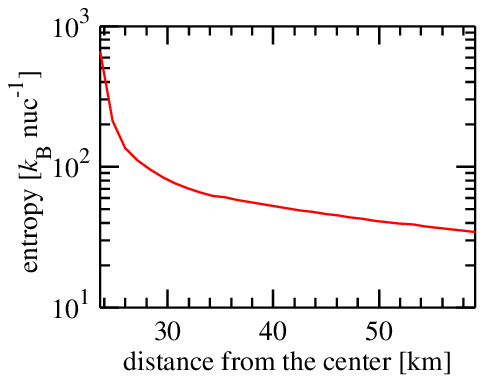}
  \includegraphics[width=0.5\textwidth]{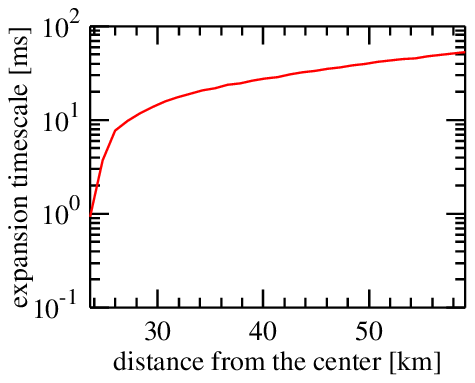}

  \caption{Left: Asymptotic entropy (at $0.5$~MeV) as a function of the
 distance from the center. Right: Expansion timescale (the $e$-folding
 time of temperature from $0.5$~MeV) as a function of the distance from
 the center.}
\end{figure}

\section{Nucleosynthesis in Winds}

The nucleosynthetic yields in each wind trajectory are obtained by
solving an extensive nuclear reaction network code. The network consists
of 6300 species between the proton and neutron drip lines, all the way
from single neutrons and protons up to the $Z = 110$ isotopes (see
\citep{Wana09} for more detail). Each nucleosynthesis calculation is
initiated when the temperature decreases to $9 \times 10^9$~K, at which
only free nucleons exist. The initial compositions are then given by the
initial electron fraction $Y_\mathrm{e0}$ (number of protons per
nucleon). In this study, $Y_\mathrm{e0}$ is taken to be a free
parameter. We explore the nucleosynthesis for all the winds with
$Y_\mathrm{e0} = 0.10, 0.15, 0.20, 0.25, 0.30$, which are consistent
with a recent hydrodynamic study of BH-NS \citep{Surm08}. Note that the
initial $Y_\mathrm{e}$ in the torus, consisting of decompressed NS
matter is low, and $Y_\mathrm{e}$ in the outgoing wind remains to be low
because $L_{\bar{\nu}_e} > L_{\nu_e}$ for the torus during a significant
time of its evolution (e.g., \citep{Seti06}).

The neutron-to-seed ratios at the onset of $r$-processing (defined at
$2.5 \times 10^9$~K) are shown in Figure~4 (left panel). Note that
$Y_\mathrm{e}$ at this stage is $\sim 0.1$ higher than $Y_\mathrm{e0}$
owing to the neutrino effects, which is obviously overestimated in our
assumption of $L_{\bar{\nu}_e} = L_{\nu_e}$. In all the $Y_\mathrm{e0}$
cases, the neutron-to-seed ratios are substantially higher than 100
(that is required for the 3rd $r$-process peak formation) in the
innermost winds owing to the high entropies and the short expansion
timescales (Figure~3), where the fission cycling is expected. In the
outer winds, however, only the low $Y_\mathrm{e0}$ case attains a high
neutron-to-seed ratio (up to $\sim 70$) because of the moderate
entropies and expansion timescales.

For each $Y_\mathrm{e0}$ case, the nucleosynthetic yields are
mass-averaged over the entire range of $R_\nu$ (from $2 R_\mathrm{S}$ to
$5 R_\mathrm{S}$), which is shown in Figure~4 (right panel). Despite the
high neutron-to-seed ratios in the inner winds, the $Y_\mathrm{e0} =
0.25$ and 0.30 cases contribute only up to the 2nd $r$-process peak ($A
= 130$) because of the very small $\dot{M}$ in the inner winds
(Figure~2, right panel). Our result indicates that neutron-rich winds
with $Y_\mathrm{e0} < 0.20$ ($< 0.30$ at the onset of $r$-processing)
are required to account for the 3rd $r$-process peak formation ($A =
195$). Notable is that the ``envelope'' made by the curves for various
$Y_\mathrm{e0}$ reasonably fits the solar $r$-process distribution. This
implies that the wide range of $Y_\mathrm{e}$ (in terms of space and
time) in the presented case leads to production of all the heavy
$r$-process nuclei.

\begin{figure}[t]
  \includegraphics[width=0.5\textwidth]{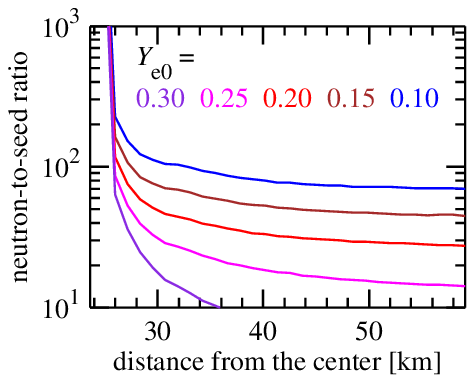}
  \includegraphics[width=0.5\textwidth]{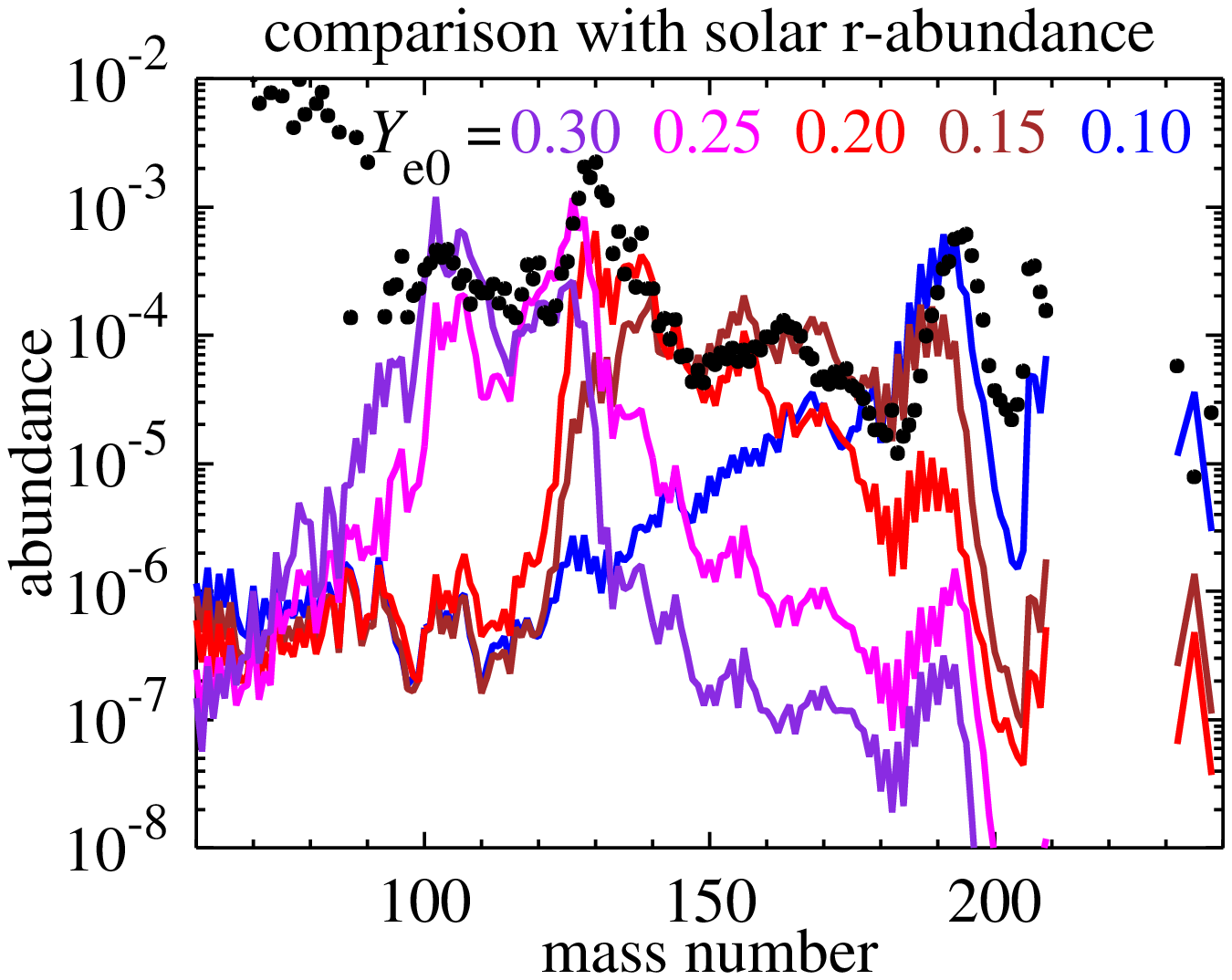}

  \caption{Left: Neutron-to-seed ratios at the onset of $r$-processing
 ($2.5\times 10^9$~K) as a function of the distance from the center for
 various initial electron fractions ($Y_\mathrm{e0} = 0.10$, 0.15, 0.20,
 0.25, and 0.30). Right: Mass-averaged nucleosynthetic yields for
 various initial electron fractions (lines), which are compared with the solar
 $r$-process distribution (dots).}
\end{figure}

\section{Implications}

Our model of black hole winds suggests that the innermost wind
trajectories attain substantially higher entropies ($> 100 k_\mathrm{B}$
per nucleon) and shorter expansion timescales ($< 10$~ms). This
indicates that all the relevant astrophysical conditions, i.e., NS-NS
and BH-NS mergers and collapsars (or hypernovae) are potential factories
of the $r$-process nuclei. However, our nucleosynthesis result shows
that significant neutron-richness in the wind is still required in order
to account for the formation of the 3rd $r$-process peak. In this
regard, NS-NS and BH-NS are favored compared to collapsars, since the
accretion tori originate from neutron-star matter (and moreover,
$L_{\bar{\nu}_e} > L_{\nu_e}$) in the former case and iron-peak (or
alpha) elements in the latter, respectively.

Obviously, more elaborate hydrodynamical studies of the relevant
astrophysical sites are needed to obtain information of the neutrino
field that controls the dynamics as well as the neutron-richness in the
black hole winds.  Note that the aforementioned astrophysical phenomena
are also suggested to be the sources of (short and long, respectively)
gamma-ray bursts. An interesting possibility in this context is that the
neutron-rich nuclei ejected in and after NS-NS or BH-NS mergers might
lead to detectable transient electromagnetic signal \citep{Metz10}. The
studies of Galactic chemical evolution will be also important to test
the contributions of NS-NS, BH-NS, and collapsars (or hypernovae) to the
enrichment history of the $r$-process elements in the Milky Way.

\begin{theacknowledgments}
The project was supported by the Deutsche Forschungsgemeinschaft through
Cluster of Excellence EXC~153 ``Origin and Structure of the Universe''
(http://www.universe-cluster.de).
\end{theacknowledgments}

\end{document}